\documentclass[
 amsmath,amssymb,
 aps,
prb,twocolumn, superscriptaddress]{revtex4-2}

\usepackage{graphicx}
\usepackage{dcolumn}
\usepackage{bm}
\usepackage{xcolor}
\usepackage[colorlinks= true,urlcolor=blue,linkcolor= blue,citecolor=blue,bookmarks=false,pdfstartview=]{hyperref}
\usepackage{amsmath}
\usepackage{cleveref}
\usepackage{physics}
\usepackage[utf8]{inputenc}
\usepackage[T1]{fontenc}
\usepackage{verbatim}
\usepackage{textcomp}
\usepackage{textcomp}
\usepackage{graphicx}
\usepackage[utf8]{inputenc}

\usepackage{hyperref}

\begin{document}

\title{Inverse Melting of 3D Antiferromagnetic Order in Multi-sublattice Magnetic Perovskites}

\author{Bangye Qin}
\email{bangye.qin.19@ucl.ac.uk}
\affiliation{Department of Physics and Astronomy, University College London,
Gower Street, London WC1E 6BT, United Kingdom}

\author{Dmitry D. Khalyavin}
\affiliation{ISIS Facility, STFC Rutherford Appleton Laboratory,
Didcot, Oxfordshire OX11 0QX, United Kingdom}

\author{Ran Liu}
\affiliation{Research Center for Materials Nanoarchitectonics (MANA),
National Institute for Materials Science,
1-1 Namiki, Tsukuba, Ibaraki 305-0044, Japan}
\affiliation{Graduate School of Chemical Sciences and Engineering,
Hokkaido University,
North 10 West 8, Kita-ku, Sapporo, Hokkaido 060-0810, Japan}
\affiliation{Institute of Scientific and Industrial Research,
Osaka University,
Mihogaoka 8-1, Ibaraki, Osaka 567-0047, Japan}

\author{Kazunari Yamaura}
\affiliation{Research Center for Materials Nanoarchitectonics (MANA),
National Institute for Materials Science,
1-1 Namiki, Tsukuba, Ibaraki 305-0044, Japan}
\affiliation{Graduate School of Chemical Sciences and Engineering,
Hokkaido University,
North 10 West 8, Kita-ku, Sapporo, Hokkaido 060-0810, Japan}

\author{Alexei A. Belik}
\affiliation{Research Center for Materials Nanoarchitectonics (MANA),
National Institute for Materials Science,
1-1 Namiki, Tsukuba, Ibaraki 305-0044, Japan}

\author{Roger D. Johnson}
\affiliation{Department of Physics, Durham University,
South Road, Durham DH1 3LE, United Kingdom}
\affiliation{Department of Physics and Astronomy, University College London,
Gower Street, London WC1E 6BT, United Kingdom}

\date{\today}

\begin{abstract}
In conventional antiferromagnets a long-range ordered 3D ground state transitions to a disordered paramagnetic state on warming, often via lower dimensional spin correlations within the critical regime. Here we demonstrate a striking departure from this paradigm. Through analysis of neutron powder diffraction data, we show that the magnetic ground state of columnar-ordered quadruple perovskites, Na$R$Mn$_2$Ti$_4$O$_{12}$ ($R$ = Dy, Sm), lacks long-range order, hosting only 2D spin correlations. On warming, this disordered state transitions into a 3D long-range ordered antiferromagnetic structure prior to the phase transition to the paramagnetic state. Our results establish an unconventional order-by-heating mechanism in which intrinsic A-site chemical disorder is coupled to competing exchange interactions between the rare earth and Mn sub-lattices, leading to a novel type of magnetic phase transition.
\end{abstract}

\maketitle

Inverse melting refers to the unusual phenomenon by which an ordered phase becomes destabilized upon cooling and a more disordered state re-emerges at lower temperature \cite{Greer1988,Rastogi1999,Greer2000,Avraham2001,Shapira2018,Duhan2025,Zhang2025}. Such behavior is thermodynamically allowed when the free-energy landscape is reshaped by quenched disorder, random-field effects, or competing interactions, such that entropy is effectively carried by the ordered phase itself \cite{Greer1988,Avraham2001,Zhang2025,Schupper2004,MendozaCoto2019}. Inverse melting has been observed in a limited number of systems, including chemically disordered alloys and metallic systems \cite{Greer1988}, vortex matter in type-II superconductors \cite{Avraham2001,Duhan2025}, polymeric materials \cite{Rastogi1999}, supramolecular assemblies \cite{Shapira2018,Shapira2020}, and electronic liquid crystals \cite{Lee2014}, but it remains relatively rare in correlated magnetic materials \cite{Ye2024}. Recent studies of geometric frustration in pyrochlore antiferromagnets have demonstrated multiple instances of re-entrance in their field–temperature phase diagrams \cite{Yahne2021}, while perovskites have been shown to host reentrant spin-glasses {\cite{Laiho2001,Dho2002,Song2017}} and multi-$\mathbf{q}$ magnetic states \cite{Andriushin2024}. To broaden this phenomenology, we now consider a distinct form of re-entrant magnetic behaviour, in which intrinsic chemical disorder and competing magnetic interactions can lead to inverse melting of 3D antiferromagnetic order into a disordered ground state.

\begin{figure}[]
  \centering
  \includegraphics[width=0.5\textwidth]{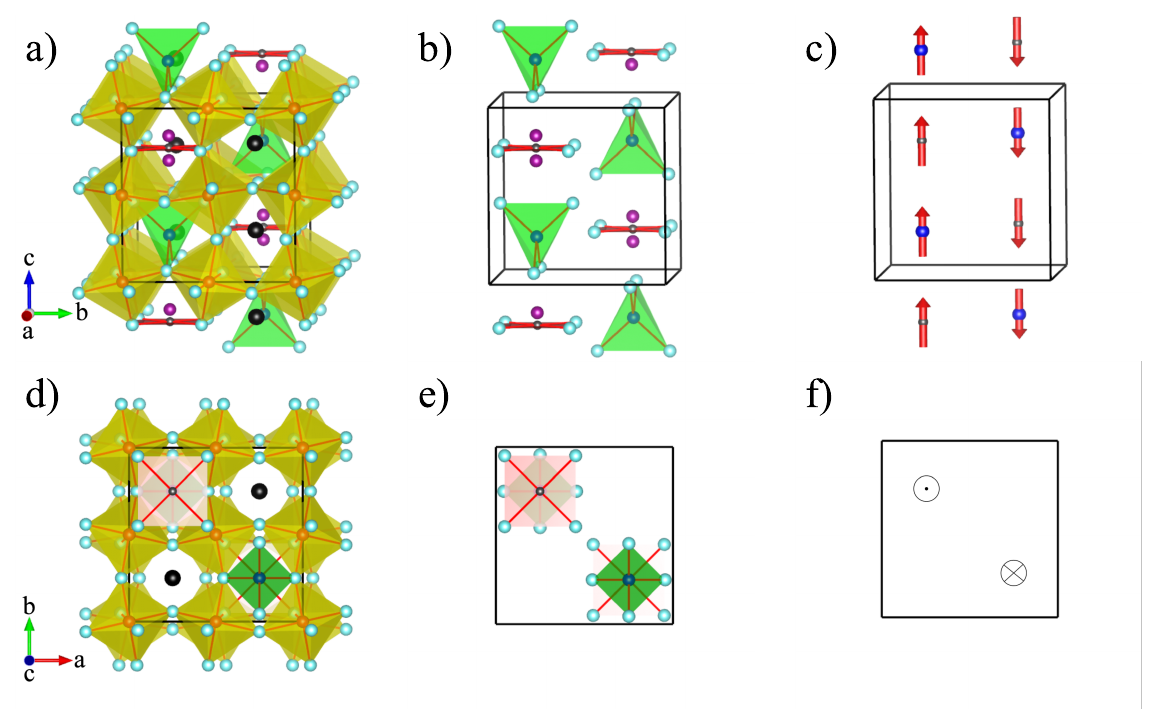}
  \caption{(a,d) The crystal structure of $\mathrm{Na}R\mathrm{Mn}_{2}\mathrm{Ti}_{4}\mathrm{O}_{12}$.  The split $\mathrm{MnO}_{4}$ square-planar units (Mn1 coordination) are shown in purple, with the ideal Mn1 positions indicated by grey spheres. The $\mathrm{MnO}_{4}$ tetrahedra (Mn2 coordination) are shown in blue, and the $\mathrm{TiO}_{6}$ octahedra in light yellow. Oxygen anions are shown in light blue, Ti ions in red, and $(\mathrm{Na}/R)$ cations as black spheres. Crystallographic axes are indicated for reference. (b,e) Isolated views of the Mn1 and Mn2 polyhedra, illustrating the alternating columnar ordering of the Mn sublattice. (c,f) The long-range ordered C-type magnetic structure, where arrows indicate the orientation of the ordered magnetic moments, and circles with dots and crosses denote moments pointing out of and into the page, respectively.
  }
  \label{fig:crystal}
\end{figure}

Quadruple perovskites  $AA'_{3}B_{4}$O$_{12}$ and their columnar variants $A_{2}A'A''B_{4}$O$_{12}$ are known to host coupled structural distortions and multiple magnetic sublattices, giving rise to competing exchange interactions and complex magnetostructural phenomena \cite{Shimakawa2008, Shimakawa2014, Chen2014,Talanov2019, Belik2018, Belik2021, Zhang2018}. Recent studies have shown that the $A_{2}A'A''B_{4}$O$_{12}$ perovskites have a unique triple-cation $A$-site ordering \cite{LiuNaRMn2020, Belik2018, Belik2018a}, stabilised by a tetragonal $a^+a^+c^-$ octahedral tilt pattern.  In such compounds, 10 coordinate $A$O$_{12}$ cages form two columns per unit cell that run parallel to the $c$-axis. Similarly, staggered columns of alternating square-planar $A'$O$_{4}$ and tetrahedral $A''$O$_{4}$ units are arranged between the $A$O$_{12}$ columns in a checkerboard pattern when viewed down $\mathbf{c}$ (see Fig.~\ref{fig:crystal}) \cite{LiuNaRMn2020, Belik2018, Belik2018a}. Further structural complexity can emerge through mixed cation occupation at the $A$-site \cite{LiuNaRMn2020}, such as in Na$R$Mn$_2$Ti$_4$O$_{12}$, where $A$ = Na$_{0.5}$R$_{0.5}$, $A'$ = Mn, $A''$ = Mn, and $B$ = Ti. These materials have been shown to develop C‑type antiferromagnetic order of the $A'A''$ Mn$^{2+}$ magnetic sublattice with $\mathbf{k}=(0,0,0)$ (Fig. \ref{fig:crystal}) immediately below $T_{\mathrm{N}}\!\sim\!8$–13\,K \cite{LiuNaRMn2020}, while a second, lower–temperature thermodynamic anomaly indicative of a low temperature phase transition at $T^*\sim6$ K appears only for selected $R$ (notably Dy, Sm, and Eu). Neutron powder diffraction experiments performed on the Dy member established C‑type magnetic order with moments $\parallel c$ below $T_{\mathrm{N}} = 12$ K, and showed a marked broadening of magnetic Bragg peaks on further cooling below $T^*=6$ K, indicative of a disordered ground state that is yet to be understood \cite{LiuNaRMn2020}. Similarly, the Sm member was found to exhibit two thermodynamic anomalies at $T_\mathrm{N}=10.5$\,K and $T^*=5$ K, but little is known about the evolution of the magnetic structure \cite{LiuNaRMn2020}.

In this paper, we present a neutron powder diffraction study of NaDyMn$_2$Ti$_4$O$_{12}$ and NaSmMn$_2$Ti$_4$O$_{12}$. The ground state of NaDyMn$_2$Ti$_4$O$_{12}$ yields broad, asymmetric magnetic diffraction intensities that are signatures of a 2D-correlated Ising-like magnetic state, here understood in terms of dense magnetic stacking faults. To the contrary, the NaSmMn$_2$Ti$_4$O$_{12}$ magnetic ground state is characterised by additional broad pseudo‑incommensurate diffraction intensities, consistent with a disordered easy‑plane modulation of Mn spins that is similarly due to dense stacking faults but under a different magnetic anisotropy. We show that on warming, both materials undergo a phase transition to a well-correlated 3D long-range ordered antiferromagnetic phase characterised by sharp magnetic diffraction intensities. The 3D ordered phase then persists up to the Neel temperature, above which both materials are paramagnetic. These remarkable results constitute a rare example of inverse-melting in an antiferromagnet, where the nature of the disordered magnetic ground state can be tuned by rare-earth single ion anisotropy.

Polycrystalline samples of Na$R$Mn$_2$Ti$_4$O$_{12}$ ($R =$ Sm and Dy) were synthesised following the high–pressure protocol reported by Liu \emph{et al.}\cite{LiuNaRMn2020}, and details are given in the Supplemental Material \cite{SM}. Neutron powder diffraction (NPD) measurements on Na$R$Mn$_2$Ti$_4$O$_{12}$ were carried out using the WISH time-of-flight diffractometer at the ISIS Neutron and Muon Source, Rutherford Appleton Laboratory (U.K.) \cite{WISH}. For the $R =$ Dy and Sm compositions, approximately 1.88~g and 2.20~g of sample, respectively,  was loaded into cylindrical 3~mm diameter vanadium cans and measured between 1.5 and 20~K. Rietveld refinements of the nuclear and magnetic structures were performed using the \textsc{fullprof} suite\cite{FullProf} against data collected in detector banks centred at average $2\theta$ values of 58$^\circ$ and 154$^\circ$.

Refinement of the NaDyMn$_2$Ti$_4$O$_{12}$ and NaSmMn$_2$Ti$_4$O$_{12}$ $P4_2/nmc$ crystal structures against neutron powder diffraction data measured at 20 and 14 K, respectively, gave an excellent fit. The structural parameters were consistent with those published in reference \cite{LiuNaRMn2020}, and are given in Table S1 of the Supplemental Material \cite{SM}. A number of weak reflections could be modeled by a minor TiO$_2$ impurity phase (2--3~wt\%) present in both samples. The refined structural models encompassed two disordered elements; 1) the $R$/Na ions randomly occupy the A-site, and 2) the square-planar coordinated Mn cation is randomly displaced above or below the coordination plane. Further, in NaSmMn$_2$Ti$_4$O$_{12}$ a small occupation of Sm at the Mn2 site was found. We note that the refined Sm:Na ratio differs from 1:1 as found by X-ray diffraction \cite{LiuNaRMn2020}. This is likely an artefact in the neutron data refinement due to strong absorption compounded by the fact that the Sm scattering length is not well known.

\begin{figure}
    \centering
    \includegraphics[width=8cm]{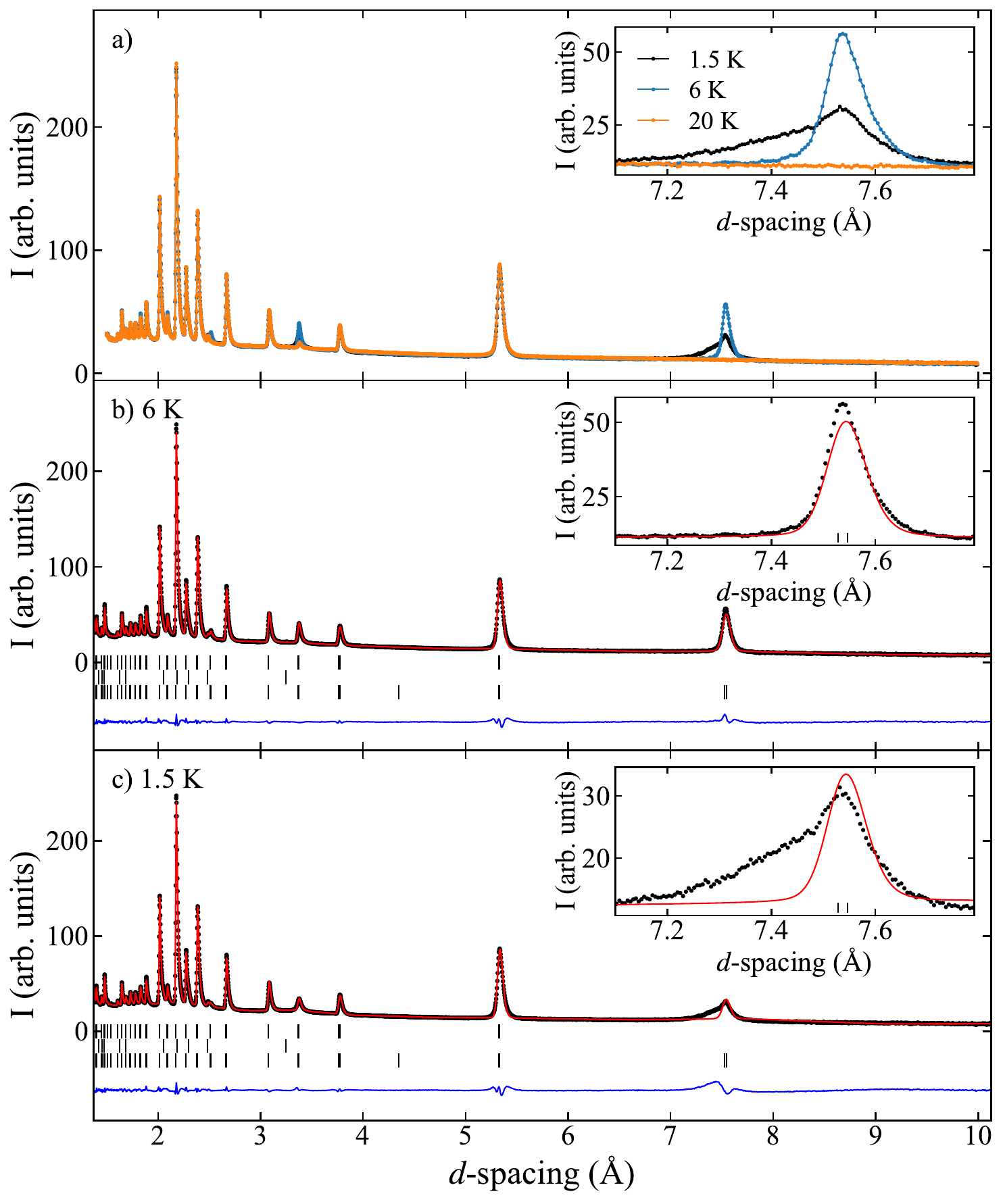}
    \caption{(a) Neutron powder diffraction (NPD) patterns of
$\mathrm{NaDyMn}_{2}\mathrm{Ti}_{4}\mathrm{O}_{12}$
measured at $T = 1.5$, 6, and 12~K.
(b) Rietveld refinement of the NPD data at $T = 6$~K,
showing the experimental data (black dots),
calculated profile (red line),
and difference curve (blue line).
The vertical tick marks indicate the Bragg reflection positions
of the nuclear phase of the main perovskite structure (top row),
the $\mathrm{TiO}_{2}$ impurity phase (middle row),
and the magnetic phase (bottom row).
(c) Neutron powder diffraction pattern at $T = 1.5$~K.
The right insets in panels (a)--(c) show an enlarged view
of the $d$-spacing region around the $\{100\}$ magnetic reflections.}
    \label{fig:Dy2}
\end{figure}

\begin{figure}[]
    \centering
    \includegraphics[width=8cm]{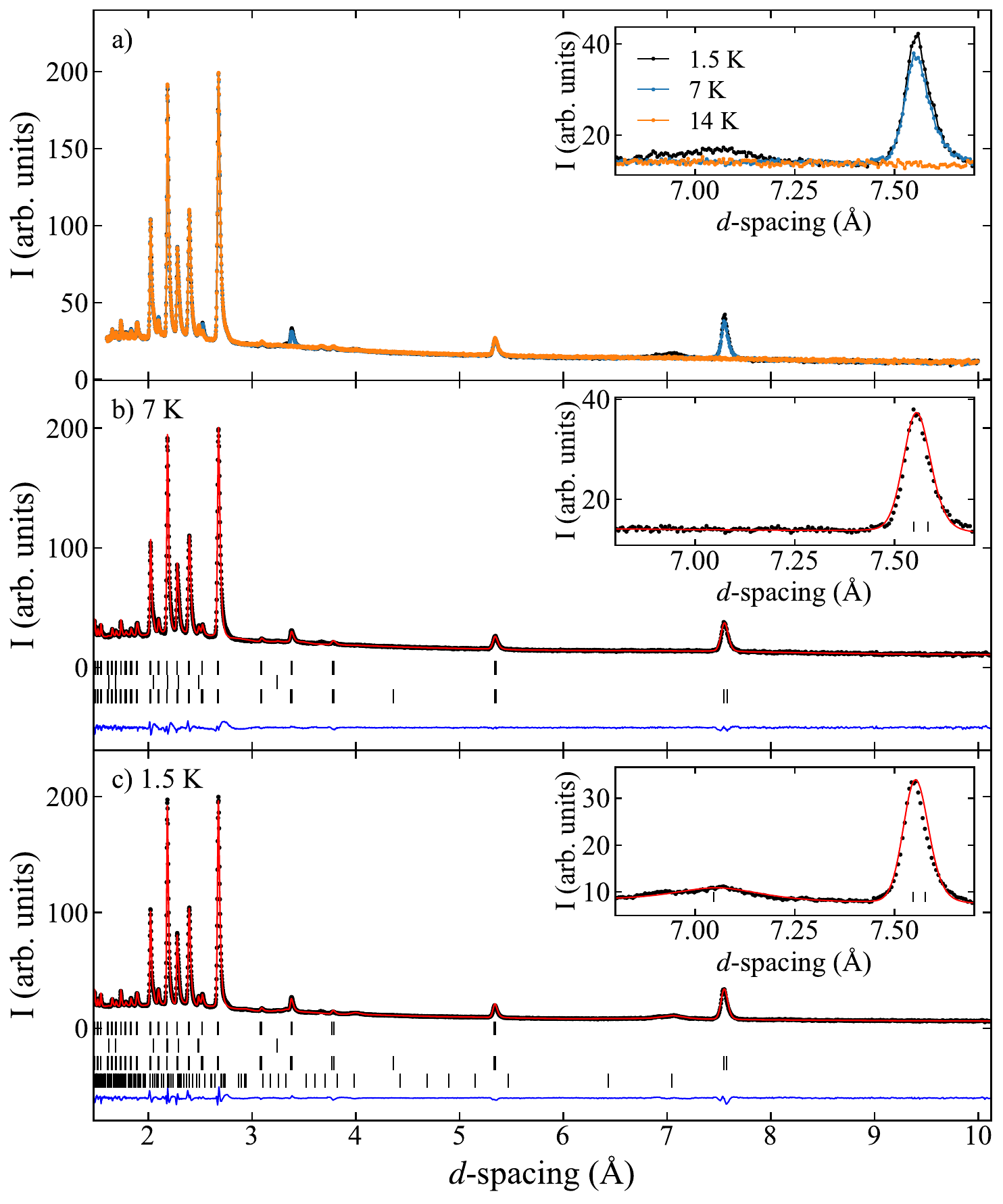}
    \caption{(a) Neutron powder diffraction (NPD) patterns of
$\mathrm{NaSmMn}_{2}\mathrm{Ti}_{4}\mathrm{O}_{12}$
measured at $T = 1.5$, 7, and 14~K.
(b) Rietveld refinement of the NPD data at $T = 7$~K,
showing the experimental data (black dots),
calculated profile (red line),
and difference curve (blue line).
The vertical tick marks indicate the Bragg reflection positions
of the nuclear phase of the main perovskite structure (top row),
the $\mathrm{TiO}_{2}$ impurity phase (middle row),
and the magnetic phase (bottom row).
(c) Neutron powder diffraction pattern at $T = 1.5$~K.
The first three rows of vertical tick marks correspond to the same
nuclear, impurity, and commensurate magnetic phases as in panel (b),
while the fourth row indicates additional magnetic reflections
associated with an incommensurate magnetic phase.
The right insets in panels (a)--(c) show an enlarged view
of the $d$-spacing region around the $\{100\}$ and $\{10\delta\}$
magnetic reflections.}
    \label{fig:Sm2}
\end{figure}

Upon cooling both materials below $T_{\mathrm{N}}$, new Bragg intensities emerge (Fig. \ref{fig:Dy2}a and Fig. \ref{fig:Sm2}a) at space group forbidden positions that index with propagation $\mathbf{k} = (0,0,0)$, signaling the onset of long-range antiferromagnetic order. The magnetic diffraction intensities could be well accounted for by a C-type magnetic structure that incorporates Mn$^{2+}$ moments on both the square-planar and tetrahedral sites. Here, antiferromagnetically coupled Mn$^{2+}$ layers in the $ab$ plane are stacked ferromagnetically along $c$, and the appearance of magnetic intensity at (100), together with the absence of (001), uniquely determines the ordered moment direction along the tetragonal $c$ axis \cite{LiuNaRMn2020}. Rietveld refinement of this magnetic structure against the neutron powder diffraction data measured at 6 K yields an excellent fit as shown in Figs. \ref{fig:Dy2}b and Fig. \ref{fig:Sm2}b.

\begin{figure}
    \centering
    \includegraphics[width=8.0cm]{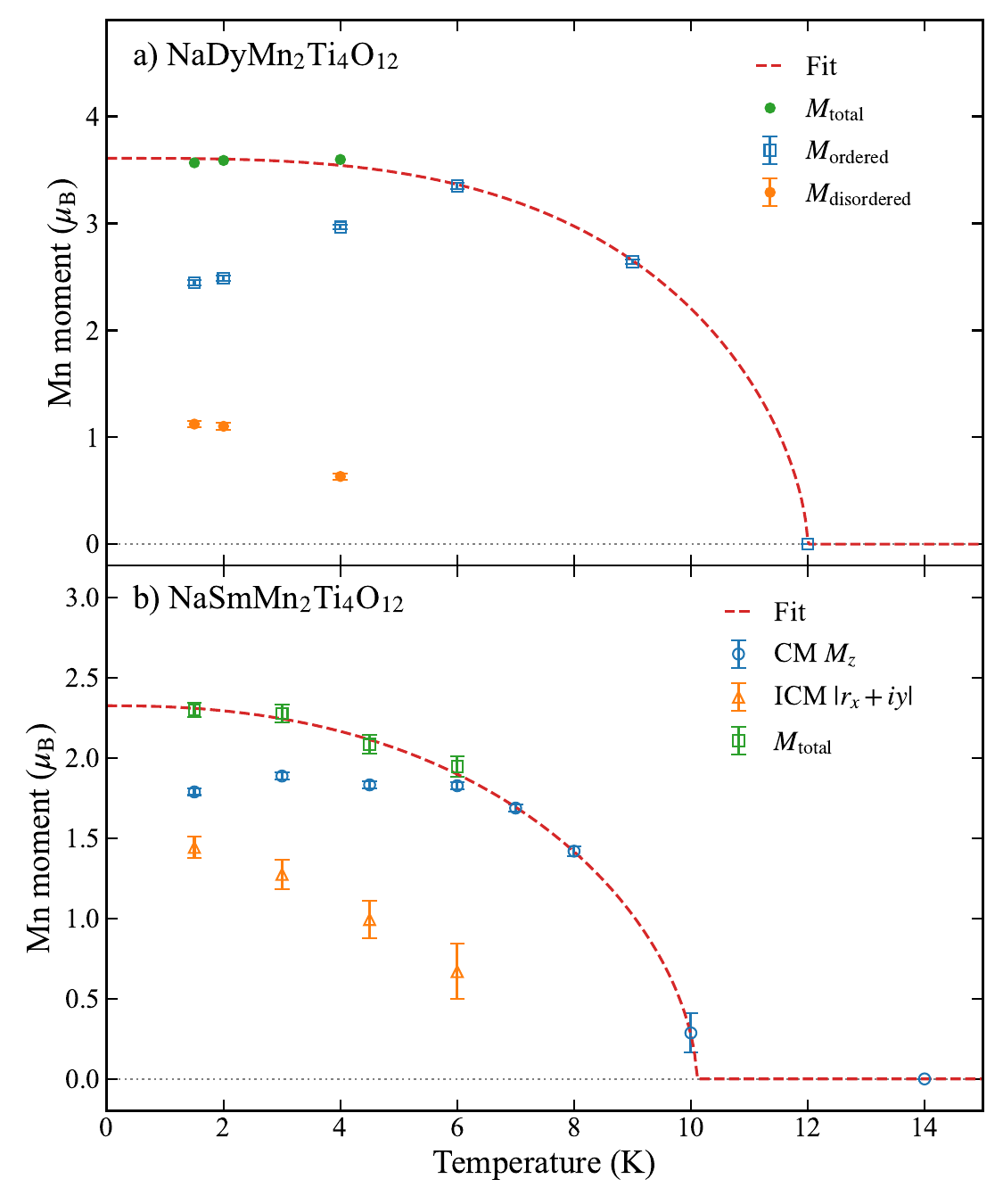}
    \caption{Temperature dependence of the Mn magnetic moment in
$\mathrm{Na}R\mathrm{Mn}_{2}\mathrm{Ti}_{4}\mathrm{O}_{12}$.
(a) $\mathrm{NaDyMn}_{2}\mathrm{Ti}_{4}\mathrm{O}_{12}$, in which
Ising-like commensurate Mn spin order (blue squares) and Mn spin disorder (orange circles) are observed.
(b) $\mathrm{NaSmMn}_{2}\mathrm{Ti}_{4}\mathrm{O}_{12}$, in which orthogonal commensurate (blue circles) and pseudo-incommensurate (oranghe triangles) Mn spin components are observed. The dashed lines represent fits of a phenomenological power-law expression to the total moment.}
    \label{fig:Mn}
\end{figure}

While the magnetic diffraction patterns immediately below $T_{\mathrm{N}}$ are resolution-limited, below $T^*\sim6$ K the $\{100\}$ peak from NaDyMn$_2$Ti$_4$O$_{12}$ develops a severe asymmetric broadening on cooling into the ground state (see Fig.~\ref{fig:Dy2}a). Nuclear reflections remain sharp, excluding instrumental or structural origins. The broad, asymmetric peak reflects a temperature-dependent reduction in the magnetic correlations along the $c$ axis, indicating that the antiferromagnetic C-type order becomes destabilized into a faulted layered state at the lowest temperatures. To estimate the evolution of the ordered component, we refined the above magnetic structure model against data measured at 4, 2 and 1.5~K. Fig.~\ref{fig:Dy2}c shows the fit to NPD data measured at 1.5~K, showing excellent agreement throughout most of the pattern, but failing to account for the broadened magnetic intensity, as expected (see Fig.~\ref{fig:Dy2}c inset). The temperature dependence of the ordered Mn moment is shown as blue circles in Fig.~\ref{fig:Mn}a.

Different behaviour was observed in NaSmMn$_2$Ti$_4$O$_{12}$. On cooling below $T^*$, the sharp resolution–limited antiferromagnetic intensities persist, while additional broad magnetic intensities at incommensurate satellite positions emerge (see Fig.~\ref{fig:Sm2}c and inset to Fig.~\ref{fig:Sm2}a). These incommensurate intensities index with wavevector $\mathbf{k}=(0,0,0.385)$. A magnetic structure model including the C–type magnetic structure with moments $\parallel$ $c$, augmented by an incommensurate helix with moments $\perp$ $c$ (a conical structure), was refined against NPD data measured at 1.5 K (Fig.~\ref{fig:Sm2}a). A severely broadened peak profile function was used for the incommensurate component, which points towards a \emph{pseudo}-incommensurate modulation of the C-type magnetic structure discussed later. Repeat refinements against data measured at 2 and 4 K were used to quantify the thermal evolution of both the well-ordered commensurate C–type and the broad incommensurate magnetic structure components. The reduction in the ordered component is observed below $T^*$, concomitant with the evolution of the incommensurate component. The total Mn moment was calculated and plotted below $T^*$ (Fig.~\ref{fig:Mn}b), which accurately follows a phenomenological fit using the expression $M_\mathrm{Mn}(T) = M_0\left(1 - \left(\frac{T}{T_{N}}\right)^\alpha\right)^\beta$, where $\alpha$ and $\beta$ are free parameters. Together, these results show that both NaSmMn$_2$Ti$_4$O$_{12}$ and NaDyMn$_2$Ti$_4$O$_{12}$ share the same tetragonal crystallographic framework and a disordered ground state, reached by inverse melting of an intermediate 3D ordered C–type antiferromagnetic structure. However, the exact nature of the disordered ground state differs between these two materials, indicating that the choice of rare-earth ion is key. Indeed, in NaYMn$_2$Ti$_4$O$_{12}$ the C-type Mn ordered moment exhibits a conventional order-parameter behavior at all temperatures below $T_\mathrm{N}$  \cite{Scatena2023}, highlighting the essential role of $f$–$d$ interactions in this material system.

To establish greater insight into the two magnetic ground states we simulate the elastic powder averaged neutron scattering intensity close to the $\{100\}$ magnetic Bragg reflection using a minimal magnetic structure model (see Supplemental Material \cite{SM} for calculation details). We construct a {$200\times200\times2000$} crystallographic supercell, in which each unit cell consists of two Mn--$R$ layers stacked along the crystallographic $c$ axis. Within each layer, the Mn$^{2+}$ sublattice comprises two crystallographically inequivalent sites associated with square-planar (S-Mn) and tetrahedral (T-Mn) coordinations. These two Mn sublattices are displaced by a $[0.5,0.5,0]$ translation and are antiferromagnetically aligned (see Fig. \ref{fig:crystal}). In adjacent layers the in-plane positions of the S-Mn and T-Mn sites are exchanged by symmetry, and so too are the relative signs of the magnetic moments. Hence, in the ordered state nearest neighbour moments separated along the crystallographic $c$ axis are ferromagnetically aligned between successive layers. Magnetic disorder is then introduced as a layer-dependent flip in magnetic moment components that occurs with probability, $p$. Hence, the ferromagnetic stacking sequence along $c$ incurs faults, while each individual layer remains internally coherent.

\begin{figure}
    \centering
    \includegraphics[width=8.0cm]{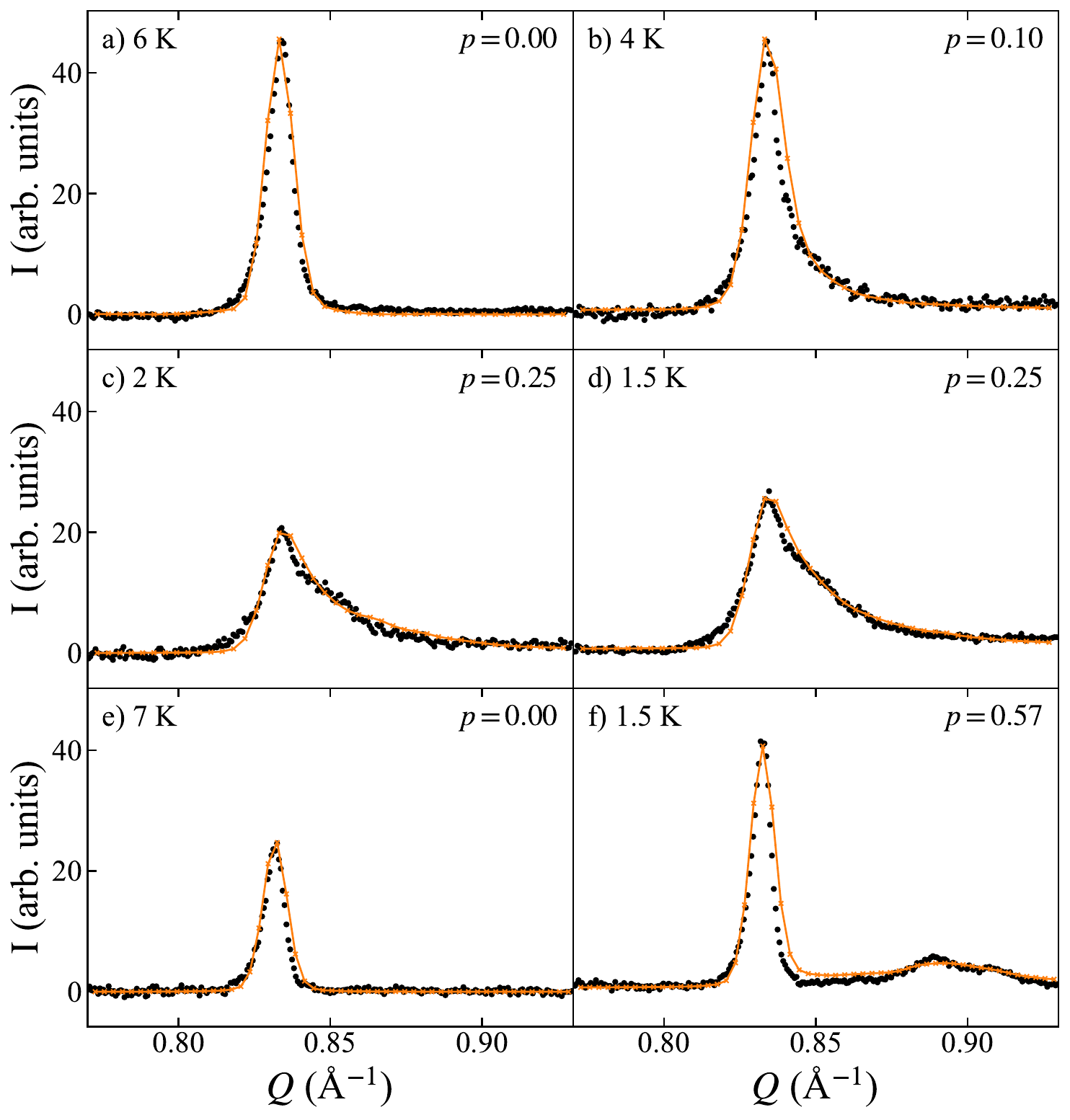}
    \caption{
Comparison between experimental and calculated scattering patterns for the $\{100\}$ magnetic Bragg peak of $\mathrm{Na}R\mathrm{Mn}_{2}\mathrm{Ti}_{4}\mathrm{O}_{12}$ ($R=\mathrm{Dy}, \mathrm{Sm}$).
(a-d) The $\{100\}$ magnetic Bragg peak from $\mathrm{NaDyMn}_{2}\mathrm{Ti}_{4}\mathrm{O}_{12}$ measured on cooling. (e,f) The $\{100\}$ magnetic Bragg peak from $\mathrm{NaSmMn}_{2}\mathrm{Ti}_{4}\mathrm{O}_{12}$ at 7~K and 1.5~K.
The experimental neutron powder diffraction data are shown as black dots, and the calculated scattering patterns (see text for details) are shown as orange curves. The magnetic stacking-fault probability, $p$, is indicated in each panel.}
    \label{fig:Model}
\end{figure}

For $\mathrm{NaDyMn}_{2}\mathrm{Ti}_{4}\mathrm{O}_{12}$, all magnetic moments were constrained to be aligned (anti)parallel to the $c$-axis, and a minimum 2-layer domain size was imposed, discussed later. Fig.~\ref{fig:Model} shows the results of these calculations, where $p$ has been refined against the diffraction data measured on warming. This Ising-like magnetic stacking fault model well reproduces the experimental data, with the probability of a fault being maximal (25\%) in the ground state and decreasing to zero on warming as $T^*$ is approached. The square root of the simulated $\{100\}$ integrated intensity is plotted in Fig.~\ref{fig:Mn}a, having been scaled to the refined ordered moment at 6 K. The thermal evolution of the total moment is monotonic, as expected of a conventional order parameter, while the onset of a disordered ground state below $T^*$ is clearly seen. The Sm-based compound exhibits altogether different behavior in the ground state. Motivated by the neutron powder diffraction data analysis, instead of Ising-like faults we introduce a model of Mn spin canting with stacking faults occurring in the in-plane component only. Hence, well correlated order of the Mn moment components (anti)parallel to $c$ is maintained. Again, the minimum 2-layer domain size was imposed, and the probability of a fault was refined against data measured at 1.5 and 7 K. As shown in Fig. \ref{fig:Model}, good agreement with the data is achieved for a canting angle of $15^{\circ}$ at 1.5 K, with a fault probability approaching 57\% in the ground state. Unlike a conventional incommensurate phase, which is characterized by a well-defined incommensurate propagation vector and long-range phase coherence, the present case is characterized by a pseudo-incommensurate non-stochastic faulting pattern due to the minimum domain size, which leads to the apparent shift in intensity away from the commensurate position for large values of $p$. We note that in Rietveld refinement, this scenario is indistinguishable from a poorly correlated incommensurate phase with characteristically broad diffraction peaks.

It is apparent that the choice of rare-earth ion determines the nature of Mn disorder in the ground state. The above analysis shows that the rare-earth magnetic moments do not contribute to the neutron diffraction intensities ($R$ and Na are randomly distributed throughout the crystal structure), but their effects are witnessed through faults in the Mn magnetic structure that develop below $T^*$; a temperature characteristic of the $f$-$d$ exchange energy. Dy$^{3+}$ ions typically adopt a ground state Kramers doublet with strong Ising anisotropy. In this case, the onset of long-range magnetic order of the Mn sublattice splits the Dy$^{3+}$ Kramers doublet via a finite exchange field that grows on cooling, effectively `switching-on' large local, Dy$^{3+}$ Ising moments distributed throughout the crystal. These Dy moments, in turn, impose a highly anisotropic exchange field on the Mn layers. We suggest that in Dy-rich regions, competing Dy--Dy interactions generate a sufficiently strong exchange field to destabilize the ideal interlayer ferromagnetic alignment of the C-type Mn magnetic structure, leading to Ising-like magnetic stacking faults. Similarly, the disordered magnetic ground state in $\mathrm{NaSmMn}_{2}\mathrm{Ti}_{4}\mathrm{O}_{12}$ can occurs due to $f-d$ interactions, but in the presence of easy plane magnetic anisotropy of Sm$^{3+}$, as opposed to Ising anisotropy of Dy$^{3+}$. Mean-field calculations given in the Supplemental Material show that the energetic cost of a fault decreases monotonically as the spin canting angle increases away from the Ising limit. Hence, non-Ising anisotropy permits a lower energetic barrier for fault nucleation, providing a natural mechanism for the increased probability of faulting observed in the Sm-based compound.

In summary, we have shown that the columnar-ordered quadruple perovskites Na$R$Mn$_2$Ti$_4$O$_{12}$ ($R=$ Dy, Sm) host a remarkable inverse melting of 3D long range antiferromagnetic order due to the interplay between multiple magnetic sublattices and intrinsic chemical disorder, tunable by choice of rare-earth single-ion anisotropy. This discovery marks a paradigm shift in our understanding of magnetic phase transitions in collinear antiferromagnets that is distinct from effects of topology or spin glass behaviour. The order-by-heating mechanism, through which low dimensional correlations transition into long-range 3D order on heating, is expected to be relevant across a wide range of mixed-cation perovskite-type materials, where multiple magnetic sublattices and competing interactions coexist.

\section*{Acknowledgments}\label{sec:ack}

B.Q. acknowledges institutional support from University College London. Neutron powder diffraction experiments were performed at the ISIS Neutron and Muon Source, Rutherford Appleton Laboratory, UK, under beam time proposal RB2320241. This work was partially supported by a Grant-in-Aid for Scientific Research (No. JP25K01657) from the Japan Society for the Promotion of Science.

\bibliography{refs.bib}

\end{document}